\documentclass[twocolumn,aps,pra,floatfix]{revtex4}
\usepackage{amsmath}
\usepackage{amssymb}
\usepackage{graphicx}
\usepackage{pst-all}

\newcommand{\unit}[1]{\, \textrm{#1}}

\newcommand{\ket}[1]{\lvert #1 \rangle}           
\newcommand{\bra}[1]{\langle #1 \lvert}           

\newcommand{\cbd}[4]{\sum_{k=#2}^{#1}\binom{#1}{k} \left(#3\right)^{k} \left(#4\right)^{#1-k}}  
\newcommand{\bd}[5]{\binom{#1}{#2} \left(#4\right)^{#2} \left( #5 \right)^{#3}} 
\newcommand{\betar}[3]{\frac{\Gamma(#1)}{\Gamma(#2)\Gamma(#3)}\,}

\newcommand{\esv}[0]{\langle\sigma_{z}\rangle}    
\newcommand{\eps}[0]{\varepsilon}                 
\newcommand{\p}[0]{M_{+}}                         
\newcommand{\n}[0]{M_{-}}                         
\newcommand{\Mmin}{M_\mathrm{min}}                              
\newcommand{\epp}[0]{\left(\frac{1+\eps}{2}\right)}   
\newcommand{\emp}[0]{\left(\frac{1-\eps}{2}\right)}   
\newcommand{\epn}[0]{\frac{1+\eps}{2}}                
\newcommand{\emn}[0]{\frac{1-\eps}{2}}                
\newcommand{\mpn}[0]{\tfrac{M+1}{2}}                
\newcommand{\ud}{\mathrm{d}} 
\newcommand{\U}[0]{\hat{U}_{\mathrm{alg}}}
\newcommand{\F}[1]{p_\mathrm{fail}^{\phantom{\mathrm{fail}}\mathrm{c}}(#1)}
\newcommand{\G}[2]{p_\mathrm{fail}^{\phantom{\mathrm{fail}}\mathrm{q}}(#1,#2)}
\newcommand{\pfq}[3]{p_\mathrm{fail}^{\phantom{\mathrm{fail}}\mathrm{q}}(#1,#2,#3)} 
\newcommand{\pfqbest}[2]{p_{\mathrm{fail}\, \mathrm{best}}^{\phantom{\mathrm{fail}\, \mathrm{best}}\mathrm{q}}(#1,#2)} 
\newcommand{\pfratio}[2]{p_{\mathrm{fail}\, \mathrm{ratio}}(#1,#2)} 
\newcommand{\myeqref}[1]{Eq.~(\ref{#1})}

\DeclareMathOperator{\Trace}{Tr}

\begin{document}

\author{Brandon M.~Anderson}
\affiliation{Department of Physics, University of Texas at Dallas,P.O. Box 830688, Richardson, TX 75083, USA}
\email{brandona@utdallas.edu}

\author{David Collins}
\affiliation{Department of Physics, Bucknell University, Lewisburg, PA 17837, USA}
\email{dcollins@bucknell.edu}
\thanks{Author to whom correspondence should be addressed.}

\preprint{APS/123-QED}

\title{Polarization Requirements for Ensemble Implementations of Quantum Algorithms with a Single Bit Output}

\begin{abstract}
We compare the failure probabilities of ensemble implementations of quantum algorithms which use pseudo-pure initial states, quantified by their polarization, to those of competing classical probabilistic algorithms. Specifically we consider a class algorithms which require only one bit to output the solution to problems. For large ensemble sizes, we present a general scheme to determine a critical polarization beneath which the quantum algorithm fails with greater probability than its classical competitor. We apply this to the Deutsch-Jozsa algorithm and show that the critical polarization is $86.6\%.$
\end{abstract}

\maketitle

\section{Introduction}

There are two general paradigms for implementing quantum algorithms~\cite{nielsen00}. In the first, the quantum algorithm is implemented on a single quantum system with the appropriate number of qubits and which can be prepared in a suitable pure state and is amenable to projective measurements. Most quantum algorithms are written with this in mind. In the second paradigm, the algorithm is implemented on an ensemble of of identical, non-interacting quantum computers. This is the situation with conventional room temperature, solution state NMR implementations, in which case the ensemble consists of approximately $10^{20}$ molecules ~\cite{vdsypen01,vdsypen00,marx00,cory98,chuang98a,cory97,gershenfeld97}.

In ensemble implementations each ensemble member undergoes the same unitary evolution as its companions and algorithms for the two paradigms are typically most similar in this respect. However, they differ in the initialization and measurement stages. In general an ensemble quantum computer can only be prepared in a mixed state, so that the state of any single ensemble member is not known with certainty. Also, the output from an ensemble quantum computer is an average of individual ensemble member measurement outcomes. The initialization and measurement issues have led to modifications of quantum algorithms for ensemble realizations.

The conventional approach to ensemble quantum computing initializes the ensemble in a pseudo-pure state, for which various preparation techniques have been proposed~\cite{schulman99,cory98,chuang98b,chuang98a,knill97} and which has the form 
\begin{equation}
  \hat{\rho}_i=\frac{\left(1-\eps\right)}{2^n} \hat{I}^{\otimes n}+ \eps\ket{\psi_i}\bra{\psi_i}
  \label{eq:pseudopure}
\end{equation}
where $n$ is the number of qubits, $\ket{\psi_i}$ is a known pure state and $0 \leq \eps \leq 1$ is called the \emph{polarization.} The idea is that under the collection of unitaries required to implement a quantum algorithm, $\U$, the density operator transforms to 
\begin{align}
 \hat{\rho}_\mathrm{final} & =\frac{\left(1-\eps\right)}{2^{n}}\hat{I}^{\otimes n}+ \eps \, \U \ket{\psi_{i}}\bra{\psi_{i}}\U^{\dagger} \nonumber \\
                           & =\frac{\left(1-\eps\right)}{2^{n}}\hat{I}^{\otimes n}+ \eps \, \ket{\psi_\mathrm{final}}\bra{\psi_\mathrm{final}}
 \label{eq:pseudopurefinal}
\end{align}
where
\begin{equation}
  \ket{\psi_\mathrm{final}} := \U \ket{\psi_{i}} 
\end{equation}
and this is followed by measuring the expectation value of a traceless observable. The identity component of of $\hat{\rho}_\mathrm{final}$ does not contribute to this measurement outcome and it is as though the pure state algorithm represented by $\ket{\psi_{i}} \rightarrow \U \ket{\psi_{i}}$ has been implemented.

Much of the discussion of ensemble quantum computing on pseudo-pure states has focused on the scaling properties of the polarization with respect to the problem's input size~\cite{warren97} or the presence of entanglement in these~\cite{schack99}. In particular, most pseudo-pure state preparation schemes result in polarizations which diminish exponentially as the number of qubits increases, thus resulting in exponentially decreasing output signal strength. However, a promising new approach using NMR with parahydrogen induced polarization attains high polarizations and appears to avoid these problems~\cite{anwar04}.   

Here we consider how well an ensemble quantum algorithm, for a given polarization and ensemble size, performs in relation to competing classical probabilistic algorithms. We propose a criterion, considering the ensemble size as one of the resources, for which an ensemble algorithm can be compared fairly to a classical competitor. We then use this to ask, for a certain class of problems, whether there is a critical polarization below which the quantum algorithm fails with greater probability than the classical algorithm.      

The remainder of this paper is organized as follows. In section ~\ref{sec:general} we provide a general scheme for comparing the performance of ensemble quantum algorithms to their classical counterparts. We only consider algorithms for which the output is obtained after measuring a \emph{single qubit.} In section ~\ref{sec:dj} we apply the general scheme to the Deutsch-Jozsa algorithm determine the critical polarization below which the quantum algorithm fails with greater probability than a classical random algorithm. Finally, the appendices contain much of the mathematical derivations of various essential results.

\section{Performance of Ensemble Quantum Algorithms vs.\ Classical Probabilistic Algorithms}
\label{sec:general}

We consider problems which take one of many possible inputs and determine into which of \emph{two possible classes} the input falls. Any classical algorithm to solve one of these could be designed to write the output to one bit; those inputs returning ``0'' fall into ``class 0'', and those returning ``1'' fall into ``case 1.'' We assume that a quantum algorithm exists, which, when applied to a collection of qubits in an appropriate pure initial state, can determine the input class with certainty. It is convenient to split the collection of qubits into a single qubit target register, on which a measurement will reveal the input type, and a remaining $n$-qubit argument and workspace register as may be required by the algorithm. This quantum analog proceeds as:
\begin{equation}
\ket{\psi_{i}}\stackrel{\U}{\longrightarrow}\left\{
\begin{array}{ll}
\ket{\phi_{0}}_\mathrm{a}\ket{0}_\mathrm{t} & \textrm{for ``class 0''}\\
\ket{\phi_{1}}_\mathrm{a}\ket{1}_\mathrm{t} & \textrm{for ``class 1''}
\end{array}\right.
 \label{eq:finalstate}
\end{equation}
where the subscripts denote the argument/workspace and target registers and $\ket{\phi_{0}}_\mathrm{a}$ and $\ket{\phi_{1}}_\mathrm{a}$ are normalized but not necessarily orthogonal argument register states. The input class is revealed following  a computational basis measurement on the target qubit.

On an ensemble quantum computer initially in the pseudo-pure state of \myeqref{eq:pseudopure}, the typical protocol~\cite{knill98,chuang98a} for determining the input class is based on the expectation value for the target qubit,
\begin{equation}
\esv_\mathrm{t} =\left\{
  \begin{array}{ll}
    \eps & \textrm{for ``class 0''}\\
    -\eps & \textrm{for ``class 1''.}
  \end{array} \right.
\label{eq:esv}
\end{equation}
This evidently allows one to distinguish the input class by ``measuring an expectation value'' (provided that the polarization is suitably large for detection in a particular experimental setup) and checking whether it is $+\eps$ or $-\eps.$  However, for an ensemble with a finite number of members $M$ and whose final state is mixed as in \myeqref{eq:finalstate}, the random nature of the target qubit measurement outcomes on individual ensemble members generates statistical fluctuations which will yield outcomes that are almost never precisely $\esv_\mathrm{t} = \pm \eps$. It is then essential to elaborate the protocol for deciding the input class, determine the probability with which this gives a correct result and compare this to a classical probabilistic algorithm which uses the same resources.

The protocol which we advocate replaces $\esv_\mathrm{t}$ by a suitable sample average of computational basis measurement outcomes over all the ensemble members. We assume that a computational basis measurement is performed on each ensemble member and that each measurement outcome is scaled to be compatible with the eigenvalues of $\sigma_{z}$, i.e.\ let $z_{j}=+1$, $z_{j}=-1$ correspond to the outcome of the measurements associated with projectors $\hat{P}_{0}=\ket{0}\bra{0}$ and $\hat{P}_{1}=\ket{1}\bra{1}$ respectively. These yield a sample average
\begin{equation}
 \bar{z}:=\frac{1}{M}\sum_{i=1}^{M}z_{i}
 \label{eq:sampave}
\end{equation}
which typically approximates $\esv_\mathrm{t}$ well as $M \rightarrow \infty.$ This leads to the decision protocol:
\begin{equation}
  \renewcommand{\arraystretch}{1.25}
  \begin{array}{ll}
     \bar{z} > 0 \; \Rightarrow \;  & \textrm{input is ``class 0'',}\\
     \bar{z} = 0 \; \Rightarrow \;  & \textrm{\parbox{2.25in}{guess the input class with probability $1/2$ for either type, and}}\\
     \bar{z} < 0 \; \Rightarrow \;  & \textrm{input is ``class 1''.}\\
  \end{array}
\label{eq:sampprotocol}
\end{equation}
This amounts a majority vote on the number of individual ensemble member outcomes which are $z_{j}=+1$ or $z_{j}=-1$ or a completely unbiased guess whenever the numbers of the two outcomes are identical. Let $\p$ be the number of times that that $z_{i}=+1$ and $\n$ the number of times that $z_{i}=-1$. It is straightforward to verify that 
\begin{equation*}
 \bar{z}:= \frac{\Delta M}{M}
\end{equation*}
where $\Delta M := \p - \n$ represents the excess of positive measurement outcomes. The protocol of \myeqref{eq:sampprotocol} assumes the best possible resolution in the measuring apparatus. That is, one can distinguish between $\Delta M = \pm 1$ (for $M$ odd) or $\Delta M = -2,0,$ or $2$ (for $M$ even). We refer to this as the \emph{best resolution case.} We shall later generalize this to arbitrary measurement resolution and demonstrate that the best resolution case is optimal.

The probability with which the quantum algorithm misidentifies the input type can be determined by considering the various routes to failure. The probability that that a ``class 0'' input will be misidentified as ``class 1''  will be denoted as $p_{\mathrm{fail}\, \mathrm{best}\, 0}$ and the probability that a ``class 1'' input will be misidentified as ``class 0'' as $p_{\mathrm{fail}\, \mathrm{best}\, 1}$. Assuming that an input is chosen from ``class 0'' with the same probability as from ``class 1,'' the quantum failure probability is $p_{\mathrm{fail}\, \mathrm{best}}^{\phantom{{\mathrm{fail}\, \mathrm{best}}}\mathrm{q}} = (p_{\mathrm{fail}\, \mathrm{best}\, 0} + p_{\mathrm{fail}\, \mathrm{best}\, 1})/2.$ Now suppose that the algorithm is run with a ``class 0'' input. The input will be misidentified if $\p < \n$ or if an incorrect class is guessed when $\p =\n.$ The probabilities with which these occur can be derived from those for measurement outcomes on individual ensemble members. In this case it follows from Eqs.~(\ref{eq:pseudopurefinal}) and~(\ref{eq:finalstate}) that
\begin{equation}
 \begin{split}
   \Pr(z_{i}=+1) & =\Trace \left(\hat{P}_{0}\, \hat{\rho}_\mathrm{final}\right)=\epp\\
   \Pr(z_{i}=-1) & =\Trace \left(\hat{P}_{1}\, \hat{\rho}_\mathrm{final}\right)=\emp.
   \label{eq:individualprobs}
 \end{split}
\end{equation}
Similarly if the algorithm is run with a ``class 1'' input the failure probability can be determined by switching $\p$ with $\n$ in the conditions for misidentification and $z_{i}=+1$ with $z_{i}=-1$ in \myeqref{eq:individualprobs}. The symmetry in these situations implies that $p_{\mathrm{fail}\, \mathrm{best}\, 1} = p_{\mathrm{fail}\, \mathrm{best}\, 0}$ and thus $p_{\mathrm{fail}\, \mathrm{best}}^{\phantom{{\mathrm{fail}\, \mathrm{best}}}\mathrm{q}} = p_{\mathrm{fail}\, \mathrm{best}\, 0}.$ Since measurements on each ensemble member amount to a Bernoulli trial the ``class 0'' failure probability is a cumulative binomial distribution. The precise form of this depends on whether $M$ is even or odd. For odd $M$, the case $\p =\n$ cannot occur and 
%
\begin{align}
 \pfqbest{\eps}{M} & =  \Pr(\n>\p)\nonumber\\
             & =  \Pr(\n \geq \tfrac{M+1}{2})\nonumber\\
             & =  \cbd{M}{{\tfrac{M+1}{2}}}{\emn}{\epn},
\label{eq:pfailodd}
\end{align}
%
indicating the dependence of the failure probability on polarization and ensemble size. For even M, the case $\p =\n$ can occur and
%
\begin{align}
 \pfqbest{\eps}{M} & = \Pr(\n>\p) + \frac{1}{2} \Pr(\n = \p)\nonumber\\
             & =  \Pr(\n \geq \tfrac{M}{2} +1) + \frac{1}{2} \Pr(\n = \tfrac{M}{2}) \nonumber\\
             & =  \cbd{M}{\tfrac{M}{2}+1}{\emn}{\epn}\nonumber\\
             & \quad + \frac{1}{2}\, \bd{M}{M/2}{M/2}{\emn}{\epn}.
\label{eq:pfaileven}
\end{align}
%

The best resolution case assumes that the measurement apparatus allows one to distinguish between two circumstances where the values of $\Delta M$ differ by as little as $2$ and thus values of $\bar{z}$ which differ by as little as $2/M.$ In a \emph{general resolution case} we assume that one can only distinguish between two situations where the values of $\Delta M$ differ by a \emph{resolution} of at least $R,$ which could depend on $M.$ In the context of the protocol of \myeqref{eq:sampprotocol} this means that outcomes for which $ -R/2 < \Delta M < R/2$ can be regarded as pure noise. The maximum magnitude of the sample average associated with this noise is $\lvert \bar{z} \rvert = R/2M$ and noting that the maximum sample average associated with any outcome has magnitude $\lvert \bar{z} \rvert = 1$, the signal to noise ratio is represented by $R/2M.$ This can be used as a guide to precise behavior of the resolution as a function of ensemble size, which may depend on the details of the apparatus. Regardless of these details, the decision protocol for the general resolution case is:
\begin{equation}
  \renewcommand{\arraystretch}{1.25}
  \begin{array}{rll}
     & \bar{z} \geq \frac{R}{2M} \; & \Rightarrow \;  \textrm{input is ``class 0'',}\\
     \frac{R}{2M} > & \bar{z} > - \frac{R}{2M} \; & \Rightarrow \;  \textrm{\parbox{2in}{guess the input class with probability $1/2$ for either type, and}}\\
     & \bar{z} \leq - \frac{R}{2M} \; & \Rightarrow \;  \textrm{input is ``class 1''.}\\
  \end{array}
\label{eq:sampprotocolgeneral}
\end{equation}
Note that the best resolution case is represented by $R=2.$ The symmetry in this protocol again results in $p_\mathrm{fail}^{\phantom{\mathrm{fail}}\mathrm{q}} = p_\mathrm{fail 0}.$ The ``class 0'' input failure probabilities are more conveniently expressed in terms of $\n.$ To do so, note that unequivocal failure, i.e.\ $\bar{z} \leq -\frac{R}{2M},$ corresponds to $\Delta M \leq -\lceil R/2 \rceil$ and, since $2\n = M - \Delta M$ this is equivalent to $\n \geq \lceil (M + \lceil R/2 \rceil )/2 \rceil.$ For convenience define the minimum number of occurrences of $z_i=-1$ needed for unequivocal failure as
\begin{equation}
 \Mmin:=\lceil (M + \lceil R/2 \rceil )/2 \rceil.
\end{equation}
Clearly $\Mmin >M/2.$ Also, it is easily shown that the ambiguous outcome $\frac{R}{2M} > \bar{z} > - \frac{R}{2M}$ is equivalent to $ \Mmin-1 \geq \n \geq M-\Mmin +1.$ Thus the quantum algorithm fails with probability
\begin{widetext}
\begin{align}
 \pfq{\eps}{M}{\Mmin} & = \Pr(\n \geq \Mmin) + \frac{1}{2}\Pr( \Mmin -1 \geq \n \geq M-\Mmin +1)\nonumber \\
                  & = \frac{1}{2} \Bigl( \Pr(\n \geq \Mmin) + \Pr( \n \geq M-\Mmin +1) \Bigr)\nonumber \\
                  & = \frac{1}{2} \cbd{M}{\Mmin}{\emn}{\epn} + \frac{1}{2} \cbd{M}{M- \Mmin+1}{\emn}{\epn}.
\label{eq:pfailqgen}
\end{align}
\end{widetext}
Several important properties of this general quantum failure probability are proved in Appendix~\ref{app:qfp}. First, for fixed $M$ and $\Mmin$, $\pfq{\eps}{M}{\Mmin}$ is a monotonically decreasing function of $\eps$ and
\begin{align}
  \pfq{0}{M}{\Mmin} & = \frac{1}{2}\\
  \pfq{1}{M}{\Mmin} & = 0.
\end{align}
The former corresponds to a maximally mixed initial state, for which the algorithm produces a maximally mixed final state and any decisions about input classes amount to unbiased guesses. The latter case corresponds to a pure initial state, for which the algorithm never fails. Second, for fixed $\eps$ and $M$, as the resolution decreases, i.e.\ $\Mmin$ increases, $\pfq{\eps}{M}{\Mmin}$ increases. Thus the best resolution case provides a lower bound on the failure probability for the quantum algorithm, as is to be expected. This bounding property is important since it appears to be easier to arrive at certain results for the best resolution case than the general resolution case. Two important results regarding the best resolution case are also proved in Appendix~\ref{app:qfp}. First, if $M$ is odd then the best resolution case failure probabilities for $M$ and $M+1$ are equal. Second, if $M$ is odd then the best resolution failure probability for $M+2$ is strictly less than that for $M$ unless $\eps=0$ or $\eps=1$ (both statements require fixed $\eps$). Thus, in the best resolution case at least, it is advantageous to using ensembles of increasing size.

In general there are no closed form expressions for cumulative binomial distributions of the sort encountered in Eqs~(\ref{eq:pfailodd}),~(\ref{eq:pfaileven}) and~(\ref{eq:pfailqgen}). However, the following result due to Bahadur~\cite{bahadur60} can give good approximations, particularly for $M \rightarrow \infty.$ If $0< p<1$, $m$ and $n$ are positive integers, and
\begin{equation}
B_{n}(m):={\sum_{k=m}^{n}\binom{n}{k} p^{k} \left(1-p\right)^{m-k}}
\end{equation}
then, provided that $np\leqslant m\leqslant n,$
\begin{equation}
A_{n}(m)\left[1+\frac{np(1-p)}{(m-np)^{2}}\right]\leqslant B_n(m)\leqslant A_n(m)
 \label{eq:bahadur1}
\end{equation}
where
\begin{equation}
A_n(m)=\binom{n}{m} p^m \left(1-p\right)^{n-m}
       \frac{(m+1)(1-p)}{(m+1)-(n+1)p}.
 \label{eq:bahadur2}
\end{equation}
Consider first the best resolution case, in which case it is only necessary to consider situations where $M$ is odd. It is straightforward to verify that the conditions for Bahadur's approximation are satisfied for the cumulative binomial distribution of \myeqref{eq:pfailodd}. The factor on the left side of  \myeqref{eq:bahadur1} becomes
\begin{equation}
    1+\frac{np(1-p)}{(m-np)^2}  =  1+\frac{(1-\eps^2)}{(1/\sqrt{M}+\sqrt{M}\eps)^2}
    \label{eq:bahadur1limit}  
\end{equation}
and thus tends to $1$ as $M \rightarrow \infty$ provided that $\sqrt{M}\eps \rightarrow \infty$ as $M \rightarrow \infty$ (this will be shown to applicable to the Deutsch-Jozsa algorithm). In such cases the quantum error probability is well approximated by \myeqref{eq:bahadur2} after the correct substitutions for $m,n$ and $p.$
Now consider the general resolution case. Bahadur's approximation applies to the first term on the right of \myeqref{eq:pfailqgen} since $\Mmin > M/2$ but in general the conditions are not satisfied for the second term on the right of \myeqref{eq:pfailqgen}. In other cases it is shown in Appendix~\ref{app:qfp} that it applies to the second term on the right of \myeqref{eq:pfailqgen} when $\eps \geq \lceil \frac{R}{2}\rceil / M.$ Thus provided that $R$ scales as $R_0 M^\alpha$ where $R_0$ is constant and $0 \leq \alpha <1,$ the approximation applies for almost all $\eps$ as $M \rightarrow \infty.$ The result analogous to that of \myeqref{eq:bahadur1limit} must be determined for each term on the right of \myeqref{eq:pfailqgen}. For $M \gg 1$ the first term gives
\begin{equation}
    1+\frac{np(1-p)}{(m-np)^2}  =  1+\frac{(1-\eps^2)}{(\lceil R \rceil/2\sqrt{M}+\sqrt{M}\eps)^2}
    \label{eq:bahadurlimitgen1}  
\end{equation}
while for the second term it gives
\begin{equation}
    1+\frac{np(1-p)}{(m-np)^2}  =  1+\frac{(1-\eps^2)}{(1/\sqrt{M}-\lceil R \rceil/2\sqrt{M}+\sqrt{M}\eps)^2}
    \label{eq:bahadurlimitgen2}  
\end{equation}
Again, these tend to $1$ as  as $M \rightarrow \infty$ provided that $\sqrt{M}\eps \rightarrow \infty$ as $M \rightarrow \infty$ and the quantum failure probability is well approximated using \myeqref{eq:bahadur2} twice with appropriate $m,n$ and $p.$

It remains to compare the failure probability for a quantum algorithm to that for  competing classical probabilistic algorithms. This is easiest for algorithms, such as the Deutsch-Jozsa algorithm or search algorithms, which solve problems with the aid of an oracle. In these the input is a function $f$ drawn from one of two classes. The only aid allowed is an oracle which can evaluate $f$ at any possible argument. The task is to determine the input type with the fewest oracle queries. We henceforth restrict the discussion to such oracle query algorithms. We are concerned with cases where $M$ is very large since these are typical in NMR realizations and also the quantum failure probability in the best resolution case decreases as $M$ increases. However, the ensemble size must be included in the count of resources and we do so by incorporating this into the total number of oracle queries (this has been used in the context of ensemble realizations of the Deutsch-Jozsa algorithm on thermal equilibrium-type states\cite{arvind03}). Suppose that $\U$ invokes the oracle $q$ times. Since $\U$ is applied to each ensemble member, the aggregate number of oracle queries is $Q:=Mq.$ Thus a quantum algorithm using $q$ queries per quantum computer operating on an ensemble with $M$ members must be compared to a classical probabilistic algorithm which uses $Q$ oracle queries. Denote the classical failure probability with $Q$ oracle queries by $\F{Q}.$ It is assumed that $0 \leq \F{Q} \leq 1/2$ and that $\F{Q}$ decreases as $Q$ increases. Then the critical polarization is the minimum $\eps$ required for the quantum failure probability to drop beneath the classical failure probability, is obtained by solving $\F{Q}=\F{Mq} = \pfq{\eps}{M}{\Mmin}$ for $\eps.$ Since $\pfq{\eps}{M}{\Mmin}$ decreases monotonically from $1/2$ to $0$ with increasing $\eps,$ there will be a unique critical polarization, $\eps(M),$ for each $M$. 

The precise behavior of $\eps(M)$ depends on the behavior of the ratio the quantum failure probability to the classical failure probability as a function of $M$ as well as the behavior of the resolution as a function of $M.$ This is somewhat simplified by considering the best resolution case since it bounds the quantum failure probability for the general resolution case  from below and will provide a lower bound on $\eps(M).$ Thus consider the best resolution case. If the critical polarization is bounded from below in the sense that there exists $M_0$ and $\eps_0 >0$ such that for $M > M_0,$ $\eps(M) \geq \eps_0$ then the conditions for Bahadur's approximation apply and it gives (see Appendix~\ref{app:qfpvscfp})
\begin{equation}
  \eps(M)=\sqrt{1-\bigl[M (\F{Mq})^2\bigr]^{1/M}}
  \label{eq:epsm}
\end{equation}
for large $M.$ 

For example, consider a classical probabilistic algorithm for which $\F{Q} = 1/c^Q$ where $ c > 1.$ It is shown in Appendix~\ref{app:qfpvscfp} that if $M \geq 2/\log{c}$ then $\eps \geq \sqrt{1-1/c^2}.$ This satisfies the conditions leading to \myeqref{eq:epsm} and gives a critical polarization in the best resolution case of
\begin{equation}
  \eps(M)=\sqrt{1-\frac{1}{c^{2q}} M ^{1/M}}.
  \label{eq:exampleepsm}
\end{equation}
In the asymptotic limit, $M ^{1/M} \rightarrow 1$ as $M\rightarrow \infty$ and
\begin{equation}
  \eps(M) \rightarrow \sqrt{1-\frac{1}{c^{2q}}}.
  \label{eq:epsmasymp}
\end{equation}

The general resolution case depends on the behavior of the resolution $R$ as a function of $M.$ However, if $R$ scales as $R_0 M^\alpha$ where $R_0$ is constant and $0 \leq \alpha <1,$ then Bahadur's approximation again applies and it is straightforward to show that as $M \rightarrow \infty,$ $\Mmin \rightarrow M/2$ which approaches the best resolution case. It follows that Eqs.~(\ref{eq:epsm}) to~(\ref{eq:epsmasymp}) apply to this situation as well.

\section{Example: The Deutsch-Jozsa Algorithm}
\label{sec:dj}

The Deutsch-Jozsa problem~\cite{deutsch92} considers functions  $f:\{0,1\}^{n}\to\{0,1\}$ which are guaranteed to be either constant or balanced. A  balanced function yields $0$ for precisely half of the $N=2^n$ possible arguments and $1$ for the remaining half. The task is to identify the function type using the minimum number of invocations of an oracle which can evaluate $f(x)$ at any $x = 0,\ldots N-1.$ The approaches for determining the function type with \emph{certainty} are well-known~\cite{deutsch92,cleve98}; classically, in the worst case, the function must be evaluated for $2^{n-1}+1$ different arguments; if two different inputs have yield different outputs it is balanced but if all inputs return the same output it is constant. 

\begin{figure}[h]
 \includegraphics[scale=.9]{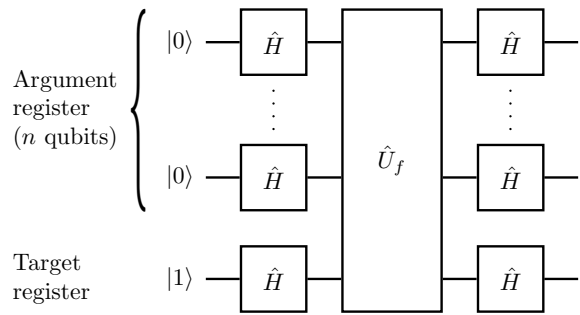} 
 \caption{Quantum circuit for the standard version of the Deutsch-Jozsa algorithm. The actions of the gates are defined in the text and the algorithm terminates with a computational basis measurement on all $n$ control register qubits.}
 \label{fig:standarddj}
\end{figure}

The circuit for the standard Deutsch-Jozsa quantum algorithm is illustrated in Fig.~\ref{fig:standarddj} where the gate operations are defined on computational basis states as
\begin{equation}
 \hat{H}\ket{x}=\frac{1}{\sqrt{2}}\sum_{y=0}^{1}\left(-1\right)^{x\cdot y}\ket{y}
\end{equation}
for the Hadamard gate and 
\begin{equation}
 \hat{U}_{f}\ket{x}\ket{y}=\ket{x}\ket{y\oplus f(x)}
\end{equation}
for the oracle. These are extended linearly to arbitrary superpositions of quantum states.
It is straightforward to demonstrate that if $f$ is constant then final state of the two registers is $\ket{\psi_\mathrm{final}} = \ket{0 \ldots 0}\ket{1}.$ while, if $f$ is balanced, $\ket{\psi_\mathrm{final}} = \sum_{x=1}^{N-1} \alpha_x \ket{x} \ket{1}.$ Notably, for a balanced function, the state $\ket{0 \ldots 0}$ does not appear in the argument register superposition. Thus an $n$-qubit computational basis measurement on the argument register reveals the function type. This quantum algorithm requires just one oracle invocation to accomplish this (giving $q=1$).

\begin{figure}[h]
 \includegraphics[scale=.9]{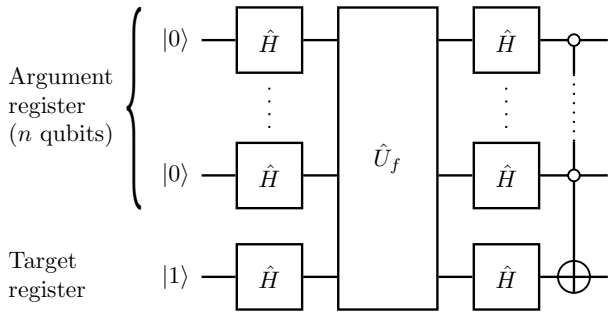} 
 \caption{Modified quantum circuit which produces a single bit output for the Deutsch-Jozsa problem. The final gate is a multiply-controlled \textsf{NOT} which applies a \textsf{NOT} to the target register when every argument register qubit is in state $\ket{0}.$}
 \label{fig:standardrevised}
\end{figure}

In the language developed earlier, the constant functions correspond to ``class 0'' and balanced functions to ``class 1'' and the algorithm should be modified so as to yield a single bit output. This is accomplished by an additional multiply-controlled \textsf{NOT} as illustrated in Fig.~\ref{fig:standardrevised}.
If $f$ is constant, the final state is of both registers is $\ket{\psi_\mathrm{final}} =\ket{\phi_0}\ket{0}$, while if $f$  is balanced, the final state will be $\ket{\psi_\mathrm{final}}=\ket{\phi_1}\ket{1}$ for some (irrelevant) $\ket{\phi_j}$.  Thus a \emph{computational basis measurement on the target qubit} reveals the function type. Note that the extra  multiply-controlled \textsf{NOT} gate can be decomposed into a sequence of $O(n^2)$ basic one and two qubit gates~\cite{barenco95}.

The framework developed earlier can be used to compare the performance of ensemble realizations of this algorithm to its classical probabilistic counterparts. The classical probabilistic algorithm proceeds by evaluating $f$ on $M<N/2+1$ distinct arguments. If all outputs are the same $f$ is identified as constant, whereas if two outputs differ $f$  will be identified as balanced. This can only fail when a balanced function happens to return the same output for all $M$ arguments. Assuming that a balanced or constant function is chosen with equal probability, it is shown in Appendix~\ref{app:cfp} that the probability with this occurs is well approximated by 
\begin{equation}
 \F{M} = \frac{1}{2^M}
\end{equation}
provided that $M \ll N/2.$ 

\begin{figure}[h]
 \includegraphics[scale=1]{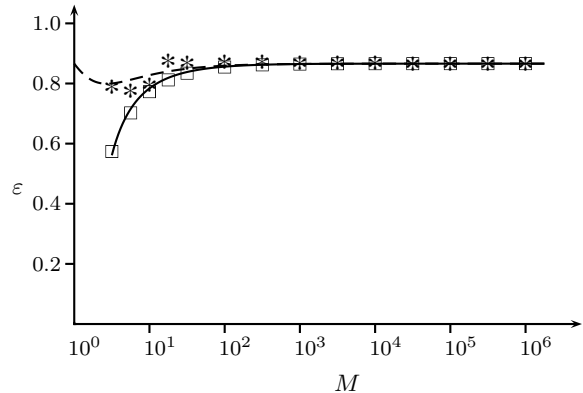}
 \caption{Critical polarization vs ensemble size for the Deutsch-Jozsa algorithm. The solid line is generated via Eq.~(\ref{eq:bestresmoderate}) while the dashed line is generated via  Eq.~(\ref{eq:bestrespol}). The squares display data obtained by solving Eq.~(\ref{eq:criticalpoldj}) numerically for the best resolution case, while the asterisks are for a resolution $R = \sqrt{M}$.}
 \label{fig:graph1}
\end{figure}

The critical polarization is determined by solving
\begin{equation}
 \G{\eps}{M} = \frac{1}{2^M}.
 \label{eq:criticalpoldj}
\end{equation}
For the best resolution case, the approximation of Eq.~(\ref{eq:epsm}) with $c=2$ gives
\begin{equation}
\eps\left(M\right)=\sqrt{1-\frac{1}{4}\,\left(M\right)^{1/M}}.
\label{eq:bestrespol}
\end{equation}
We note that a better approximation for intermediate ensemble sizes is  
\begin{equation}
\eps\left(M\right)=\sqrt{1-\frac{1}{4}\,\left(2.44 \pi M\right)^{1/M}}.
\label{eq:bestresmoderate}
\end{equation}
These are illustrated, along with data obtained by numerically solving Eq.~(\ref{eq:criticalpoldj}), in Fig.~\ref{fig:graph1}.
In the limit $M \rightarrow \infty,$ \myeqref{eq:epsmasymp} implies $\eps \rightarrow \sqrt{3/4} = 0.866025.$ By comparison a standard room-temperature, solution state NMR realization on $500\unit{MHz}$ spectrometer, using pulsed pseudo-pure preparation schemes~\cite{chuang98a,jones00,laflamme01} typically has $\eps \lesssim 10^{-5}.$ A more promising but more complicated method~\cite{anwar04} using parahydrogen induced polarization to produce a two qubit ensemble quantum computer has attained $\eps =0.9.$ It should be noted that, to date, all NMR realizations of the Deutsch-Jozsa algorithm~\cite{ermakov03,collins00,dorai00,kim00,marx00,linden98,chuang98c,jones98a} have had $n\leq5$ and $M \sim 10^{20} \gg N/2$ and, by our criteria, a classical algorithm with comparable resources would determine the function type with certainty and thus outperform these realizations.

\section{Conclusion}

In conclusion, we have provided a method for comparing the performance of ensemble versions of quantum algorithms whose output is extracted from a measurement on a single qubit to their classical probabilistic counterparts. We have applied this to realizations of the Deutsch-Jozsa algorithm and calculated the minimum polarization required for the quantum algorithm to outperform the classical probabilistic algorithm. Our calculations indicate that the standard room temperature solution state NMR approach attains polarizations several orders of magnitude too small but that newer approaches using parahydrogen induced polarization attain suitable polarizations for the ensemble quantum computer to outperform the classical probabilistic algorithm.  

\appendix

\section{Quantum failure probability}
\label{app:qfp}

The following useful representation of cumulative binomial distributions~\cite{johnson92} can be verified by repeated integration by parts:
\begin{equation}
  B_n(m):=\sum_{k=m}^{n}\binom{n}{k} p^{k} \left(1-p\right)^{ \left(m-k\right)} =I_p(m,n-m+1)
  \label{eq:cbdtoibf}
\end{equation}
where $I_p(x,y)$ is the incomplete beta function defined by:
\begin{equation}
  I_p(x,y) := \betar{x+y}{x}{y}\int_0^pt^{x-1}(1-t)^{y-1}\mathrm{d}{x}.
  \label{eq:ibfdefn}
\end{equation}

\subsection{Behavior with respect to $\eps$}

It is trivial to show by direct substitution that $\pfq{1}{M}{\Mmin} = 0.$ Now consider
%
\begin{align}
  \pfq{0}{M}{\Mmin} & = \frac{1}{2} \sum^{M}_{k=\Mmin}\binom{M}{k} \Bigl(\frac{1}{2}\Bigr)^k \Bigl(\frac{1}{2}\Bigr)^{M-k} \nonumber \\
                    & \;   + \frac{1}{2} \sum^{M}_{k=M-\Mmin+1}\binom{M}{k} \Bigl(\frac{1}{2}\Bigr)^k \Bigl(\frac{1}{2}\Bigr)^{M-k}  \nonumber \\
                    & = \frac{1}{2^{M+1}}  \sum^{M}_{k=\Mmin}\binom{M}{k} \nonumber \\
                    &  \; + \frac{1}{2^{M+1}} \sum^{M}_{k=M-\Mmin+1}\binom{M}{k} \nonumber \\
  \intertext{and, since $\Mmin > M/2,$}
    \pfq{0}{M}{\Mmin}  & = \frac{1}{2^{M}} \sum^{M}_{k=\Mmin}\binom{M}{k} \nonumber \\
                    &  \quad + \frac{1}{2^{M+1}}\sum^{\Mmin-1}_{k=M-\Mmin+1}\binom{M}{k}. \nonumber 
  \end{align}
%
It is straightforward to show that 
\[ \sum^{\Mmin-1}_{k=M-\Mmin+1}\binom{M}{k} = 2 \sum^{\Mmin-1}_{k=\lfloor (M+1)/2\rfloor}\binom{M}{k} \]
and thus
\begin{align}
  \pfq{0}{M}{\Mmin} & = \frac{1}{2^M}\, \sum^M_{k=\lfloor (M+1)/2\rfloor}\binom{M}{k} \nonumber \\
                    & = \frac{1}{2}.
\end{align}

Now consider the behavior as $\eps$ increases. We show that 
\begin{equation}
 \frac{\partial \pfq{\eps}{M}{\Mmin}}{\partial\eps} < 0
\end{equation}
for $0 < \eps < 1.$ To prove this, note that derivatives of cumulative binomial distributions are easily computed using Eqs.~(\ref{eq:cbdtoibf}) and (\ref{eq:ibfdefn});
\begin{eqnarray}
\frac{\partial B_n(m)}{\partial p} & = & \frac{\partial}{\partial p}I_p(m,n-m+1) \nonumber \\
                                   & = & \betar{n+1}{m}{n-m+1} p^{m-1}(1-p)^{n-m} \nonumber \\
                                   & > & 0 \nonumber
\end{eqnarray}
provided that $0<p<1$. Equation~(\ref{eq:pfailqgen}) shows that the quantum failure probability is just the sum of two positively weighted cumulative binomial distributions with $p=\emn$, and applying the chain rule proves the result.\\

\subsection{Behavior with respect to $\Mmin$}

We show that, for any fixed $\eps$ and $M,$
\begin{equation}
 \pfq{\eps}{M}{\Mmin+1} \geq \pfq{\eps}{M}{\Mmin}
\end{equation}
provided that $M/2 < \Mmin \leq M-1.$ Let
\begin{equation*}
  \Delta p_M:= \pfq{\eps}{M}{\Mmin+1} - \pfq{\eps}{M}{\Mmin}.
\end{equation*}
Then 
\begin{widetext}
  \begin{align}
    \Delta p_M & =
                \frac{1}{2}\, \Biggl\{ \cbd{M}{\Mmin+1}{\emn}{\epn} - \cbd{M}{\Mmin}{\emn}{\epn} \Biggr\} \nonumber \\ 
                 & \quad + \frac{1}{2}\, \Biggl\{ \cbd{M}{M-\Mmin}{\emn}{\epn} - \cbd{M}{M-\Mmin+1}{\emn}{\epn} \Biggr\} \nonumber \\ 
    \intertext{and only one term within each bracket remains, giving}
    \Delta p_M & =  - \frac{1}{2}\,\binom{M}{\Mmin} \left( \emn \right)^M \left( \epn \right)^{M-\Mmin} 
                   + \frac{1}{2}\,\binom{M}{M-\Mmin} \left( \emn \right)^{M-\Mmin} \left( \epn \right)^M  \nonumber \\ 
               & =  \frac{1}{2}\,\binom{M}{\Mmin}  \left( \emn \right)^{M-\Mmin} \left( \epn \right)^{M-\Mmin} 
                                                  \biggl\{ \left( \epn \right)^{\Mmin} - \left( \emn \right)^{\Mmin} \biggr\}. \nonumber
  \end{align}
\end{widetext}
Since $0 \leq \eps \leq 1,$ the term between brackets is positive. This proves the result.

\subsection{Best resolution case behavior with respect to $M$}

Consider the quantum failure probability for the best resolution case when $M$ is odd.  We show that the quantum failure probability remains constant if one additional ensemble member is added:
\begin{equation}
\pfqbest{\eps}{M+1}=\pfqbest{\eps}{M}.
\end{equation}
Let
\begin{equation*}
 \Delta p_M :=   \phantom{-} \pfqbest{\eps}{M+1}-\pfqbest{\eps}{M}. 
\end{equation*}
Then Eqs.~(\ref{eq:pfailodd}), (\ref{eq:pfaileven}) and~(\ref{eq:cbdtoibf}) imply, with $p:=(1-\eps)/2$, 
\begin{align}
 \Delta p_M =   & \phantom{+} \frac{1}{2}\,  I_p\bigl(\tfrac{M+1}{2},\tfrac{M+1}{2}+1 \bigr) \nonumber \\ 
                & + \frac{1}{2}\,  I_p\bigl(\tfrac{M+1}{2}+1,\tfrac{M+1}{2} \bigr) \nonumber \\ 
                & -I_p \bigl(\tfrac{M+1}{2},\tfrac{M+1}{2} \bigr)\nonumber \\ 
            =   & \phantom{+} 0 
\end{align}
since $I_p(x+1,x)+I_p(x,x+1)=2 I_p(x,x)$. This proves the result.

Now consider passing from $M$ to $M+2.$ We show that, for the best resolution case and $M$ odd,
\begin{equation}
\pfqbest{\eps}{M+2} \leq \pfqbest{\eps}{M},
\end{equation}
with equality only when $\eps =0$ or $\eps =1.$ Let
\begin{equation*}
 \Delta p_M :=   \phantom{-} \pfqbest{\eps}{M+2}-\pfqbest{\eps}{M}. 
\end{equation*}
Then Eqs.~(\ref{eq:pfailodd}) and~(\ref{eq:cbdtoibf}) imply, with $p:=(1-\eps)/2$,
\begin{widetext}
\begin{align}
  \Delta p_M(p)  & =  \phantom{+} I_p \bigl(\tfrac{M+1}{2}+1,\tfrac{M+1}{2}+1 \bigr)
                   + I_p\bigl(\tfrac{M+1}{2}+1,\tfrac{M+1}{2} \bigr) \nonumber \\
               & = \binom{M}{\mpn}\int_0^p 
                                       t^{(M-1)/2} (1-t)^{(M-1)/2}
                                       \bigl[ 2t(1-t)(M+2)-{\mpn} \bigr]\, \ud t. \nonumber
\end{align}
\end{widetext}
Now consider 
\[ g(p):=-\binom{M}{\mpn}(1-2p)p^{\mpn}(1-p)^{\mpn}.\]
It is straightforward to show that 
\[ \frac{dg}{dp} = \frac{d\Delta p_M}{dp} \] 
and that $g(0)=\Delta p_M(0).$ Since both functions are continuous it follows that they are identical,
\[ \Delta p_M = -\binom{M}{\mpn}(1-2p)p^{\mpn}(1-p)^{\mpn}.\]
However $0 < p < 1/2$ for $0 < \eps < 1,$ and thus  $\Delta p_M < 0$. For $\eps =0$ and $\eps =1,$ corresponding to $p=1/2$ and $p=0$ respectively, $\Delta p_M=0.$ This proves the result.

\section{Quantum failure probability vs classical failure probability}
\label{app:qfpvscfp}

\subsection{Applying Bahadur's approximation}

Consider a circumstance where it is known that there exist a positive integer $M_0$ and $\eps_0 >0$ such that for $M > M_0,$ $\eps(M) \geq \eps_0$ and where the resolution scales as $R_0 M^\alpha$ where $R_0 > 0$ is constant and $0 \leq \alpha <1.$ Then $\sqrt{M} \eps \rightarrow \infty$ as $M \rightarrow \infty.$ Thus the term
\begin{equation*}
    1+\frac{np(1-p)}{(m-np)^2}
\end{equation*}
in Eqs.~(\ref{eq:bahadur1limit}),~(\ref{eq:bahadurlimitgen1}) and~(\ref{eq:bahadurlimitgen2}) tends to $1$ as $M \rightarrow \infty.$ It remains to approximate $A_n(m)$ in \myeqref{eq:bahadur2}. Clearly for the resolution which scales as described above, $\Mmin \approx M/2$ if $M \gg 1$ and \myeqref{eq:pfailqgen} implies
\begin{align}
   \pfq{\eps}{M}{\Mmin} & \simeq \frac{1}{2}\, A_M(M/2) + \frac{1}{2}\,A_M((M+1)/2)\nonumber \\
                        & \simeq  A_M(M/2). \nonumber
\end{align}
For $M \gg 1$ this gives
\begin{widetext}
\begin{align}
  \pfq{\eps}{M}{\Mmin} & \simeq \binom{M}{M/2}  \emp^{M/2}\epp^{M/2}\,\frac{(M/2+1)(1+\eps)/2}{(M/2+1)-(M+1)(1-\eps)/2} \nonumber\\
                       & \simeq \binom{M}{M/2}  \frac{(1-\eps^2)^{M/2}}{2^M}\,
                                                \frac{(1+\eps)}{2 \eps}. \nonumber \\
  \label{eq:intapprox1}
\end{align}
\end{widetext}
The binomial coefficient can be approximated using Stirling's formula
\[  n!\simeq\sqrt{2\pi n}n^ne^{-n} \]
for $n \gg 1.$ Thus
\begin{align}
  \binom{M}{M/2} & = \frac{M!}{(M/2)! (M/2)!} \nonumber \\
                 & \simeq \frac{\sqrt{2\pi M} }{\pi M} \,
                     \frac{M^M}{(M/2)^M} = \sqrt{\frac{2}{\pi M}}\, 2^M, \nonumber
\end{align}
giving
\begin{equation*}
  \pfq{\eps}{M}{\Mmin} \simeq \sqrt{\frac{2}{\pi M}}\,  \frac{(1+\eps)}{2 \eps}\, (1-\eps^2)^{M/2}.
\end{equation*}
The critical polarization is determined by $ \pfq{\eps}{M}{\Mmin} = \F{Mq}$ and this yields
\begin{equation}
  1- \eps^2 = \biggl( \F{Mq} \sqrt{2 \pi M} \frac{\eps}{1+\eps} \biggr)^{2/M}.
  \label{eq:intapprox}
\end{equation}
Now consider $M \gg 1.$ Since we have assumed that $\eps(M) \geq \eps_0 \neq 0,$ the factors on the right which are constants or contain $\eps$ are approximately 1. Thus 
\begin{equation*}
  \eps^2 = 1- \biggl( \F{Mq} \sqrt{M} \biggr)^{2/M},
\end{equation*}
giving \myeqref{eq:epsm}.

Note that this can be improved for intermediate sized $M$ by retaining the factor of $\sqrt{2\pi}$ in \myeqref{eq:intapprox}. However, we found an even better approximation for the case of the Deutsch-Jozsa algorithm by solving numerically for $\eps(M)$ for small $M$ and substituting in to the last fraction on the first line of \myeqref{eq:intapprox1}. This explains \myeqref{eq:bestresmoderate}.

\subsection{Exponential classical failure probability}

Consider the case $\F{Q} = 1/c^Q$ where $ c > 1$ and $Q=Mq$ is the total number of oracle queries over the entire ensemble. We shall prove that there exist a positive integer $M_0$ and $\eps_0 >0$ such that for $M > M_0,$ $\eps(M) \geq \eps_0.$ The strategy is to consider the ratio of the best resolution case quantum failure probability to the classical failure probability,
\begin{equation*}
  \pfratio{\eps}{M} := \frac{\pfqbest{\eps}{M}}{\F{Mq}},
\end{equation*}
and to show that for some $M_0$ and $\eps_0 >0,$ $\pfratio{\eps}{M} > 1$ when $M>M_0$ and $\eps_0 > \eps.$ This establishes that the critical polarization, for which $\pfratio{\eps}{M}=1,$ is bounded from below by $\eps_0$ and this applies regardless of the resolution, since the best resolution case provides a lower bound for polarization. 

The crux is to establish that for sufficiently small $\eps,$ $\pfratio{\eps}{M}$ increases as $M$  increases. Note that for odd $M,$ $\pfratio{\eps}{M+1} > \pfratio{\eps}{M}$ for any $\eps$ since the best resolution case quantum failure probability remains constant while the classical failure probability decreases. Thus consider   
\begin{equation*}
  \Delta p(\eps):= \pfratio{\eps}{M+2} - \pfratio{\eps}{M}
\end{equation*}
for odd $M.$ Then using \myeqref{eq:cbdtoibf} with $p:=(1-\eps)/2,$
\begin{widetext}
\begin{align}
\Delta f(\eps,M) & = c^{qM} \left[c^{2q} I_p\left(\mpn+1,\mpn+1\right)
                                        -I_p\left(\mpn,\mpn\right)     \right] \nonumber \\
                 & = c^{qM} \binom{M}{\mpn}\int_0^p(t(1-t))^{(M-1)/2}\, [2t(t-1)(M+2)c^{2q}-\mpn]\,\ud t.\nonumber 
\end{align}
Then
\begin{equation*}
\frac{\partial \Delta f}{\partial\eps}= -\binom{M}{\mpn} \frac{c^{qM}}{2^M}\, (1-\eps^2)^{(M-1)/2}\bigl[(1-\eps^2)(M+2)c^{2q}-(M+1)\bigr]
\end{equation*}
\end{widetext}
which is negative if
\begin{equation}
  \eps < \sqrt{1 - \tfrac{M+1}{M+2}\, c^{-2q} }.
  \label{eq:critbound}
\end{equation}
Since the right hand side of \myeqref{eq:critbound} increases as $M$ increases, $\frac{\partial \Delta f}{\partial\eps} < 0$ when $\eps < \sqrt{1 - 2 c^{-2q}/3}$. 

Now consider $ \pfratio{0}{M} = c^{qM}/2.$ Then $\pfratio{0}{M} > 1$ when $M > M_0:=\lceil \log{2}/q \log{c} \rceil \geq 1.$ But there is some polarization $\eps^\prime>0$ such that $\pfratio{\eps^\prime}{M_0} = 1.$ Then choosing $\eps_0 = \min\{\eps^\prime, \sqrt{1 - 2 c^{-2q}/3}\} \neq 0$ (note that this is independent of $M$) implies that $\pfratio{\eps}{M} > 1$ for $M > M_0$ and $\eps < \eps_0.$ Finally this implies that $\eps(M) \geq \eps_0 \neq 0$ for $M > M_0.$


\section{Classical failure probability for the Deutsch-Jozsa problem}
\label{app:cfp}

In the classical probabilistic algorithm, $f$ is evaluated on $M \leq N/2$ arguments. The algorithm fails when a balanced function yields $M$ identical outputs. If the choice of balanced functions is unbiased, then the probability with which this occurs is the number of balanced functions for which the first $M$ arguments all return the same result divided by the total number of balanced functions. The number of balanced functions which return $0$ (or equivalently $1$) for the first $M$ arguments is $\binom{N-M}{N/2-M}$ and the total number of balanced functions is $\binom{M}{M/2}$. Thus the probability of misidentifying a balanced function is $2\,\binom{N-M}{N/2-M}\, \left/ \, \binom{N}{N/2} \right.$ where the factor of $2$ counts both the cases which output $0$ and those that output $1$. For $N \gg 2M$, this can be approximated using Stirling's formula. Thus the classical failure probability is
\begin{widetext}
\begin{eqnarray}
\binom{N-M}{N/2-M}\, \left/ \,
   \binom{N}{N/2}       \right. & = & \frac{(N-M)!(N/2)!}{N!(N/2-M)!} \nonumber \\
                                & \simeq & \sqrt{\frac{(N-M)N/2}{N(N/2-M)}}\;
                                      \left(\frac{N/2-M}{N-M}\right)^M\left(\frac{(N-M) (N/2)}{(N/2-M)N}\right)^N  \nonumber \\
& \stackrel{M \ll N/2}{\longrightarrow} & \frac{1}{2^M}.
\end{eqnarray}
\end{widetext}
Thus the classical failure probability tends to $p_\mathrm{bal}\, 2/2^M$ where $p_\mathrm{bal}$ is the probability with which a balanced function (vs a constant function) is chosen. In the unbiased case considered in this paper, $p_\mathrm{bal} =1/2.$ Thus the classical failure probability is well approximated by $1/2^M$ provided that $M \ll N/2.$

\acknowledgments

This work was supported by NSF REU grant number PHY-0097424.


\end{document}